\documentclass[11pt]{article}
\usepackage{graphicx,rotating,cite,epsfig,amsmath,amssymb}

\def\q2{$Q^2$}

\def\gev2{$~\mbox{GeV}^2$}

\font\tenrm=cmr10

 1
 1
 1
 3
 2
 2
 2

\begin{document}

\begingroup
\thispagestyle{empty}
\baselineskip=14pt
\parskip 0pt plus 5pt
\begin{center}
%{\large EUROPEAN LABORATORY FOR PARTICLE PHYSICS}
\end{center}

\begin{flushright}
%CERN--EP\,/\,2002-xxx\\
\today
\end{flushright}
\bigskip

\begin{center}
\huge \textbf{Prediction of charm-production fractions in neutrino
interactions }
\end{center}

\bigskip\bigskip
\begin{center}
{\Large G.~De Lellis, F.~Di Capua,  P.~Migliozzi} \\
 Universit\`a ``Federico II'' and INFN, Napoli, Italy
\end{center}

\begin{abstract}
The way a charm-quark fragments into a charmed hadron is a challenging
problem both for the theoretical and the experimental particle
physics. Moreover, in neutrino induced charm-production, peculiar
processes occur such as quasi-elastic and diffractive charm-production
which make the results from other experiments not directly
comparable. We present here a method to extract the charmed fractions
in neutrino induced events by using results from $e^+e^-$, $\pi N$,
$\gamma N$ experiments while taking into account the peculiarities of
charm-production in neutrino interactions. As results, we predict the
fragmentation functions as a function of the neutrino energy and the
semi-muonic branching ratio, $B_\mu$, and compare them with the
available data.
\end{abstract}

\vskip 0.4cm
\vfill
\begin{center}
%(To be submitted to {\it Physics Report})
\end{center}

\endgroup

\newpage
\pagestyle{plain}
\pagenumbering{arabic}
\setcounter{page}{2}
%\tableofcontents
\newpage
\baselineskip=14pt 
\tenrm

%
%%%%%%%%%%%%%%%%%%%%%%%%%%%%%%%%%%%%%%%%%%%%%%%%%%%%%%%%%%%%%%%%%%%%%%%%%%%%%%%
% Introduction
%%%%%%%%%%%%%%%%%%%%%%%%%%%%%%%%%%%%%%%%%%%%%%%%%%%%%%%%%%%%%%%%%%%%%%%%%%%%%%%
%

\section{Introduction}
\label{introduction}

The problem of how a charm-quark fragments into a charmed hadron is
challenging both from the theoretical and the experimental point of
view. Indeed, perturbative QCD is not applicable at energies
comparable with the charm-quark mass. Therefore, parameterizations
have to be used and the parameters are determined experimentally. In
the experiments the direct identification of the charmed hadron in
the final state is only possible through the visual observation of
the hadron decay and the measurement of the kinematical variables,
practically feasible only with nuclear emulsions.

In the following, we focus on the so-called charm-production fractions
($f_h$'s); i.e.~the probability that a charm-quark fragments into a
charmed hadron $h$ ($=D^0\, , D^+\, , D_s\, , \Lambda_c$).

In this paper we review all existing data on charm-production
fractions as measured by $e^+e^-$, $\pi N$ and $\gamma N$ experiments
and predict $f_h$ in neutrino induced charm-production. Indeed, data
on charmed fractions in interactions induced by neutrinos are rather
scarce. Only one experiment, E531~\cite{Ushida:1988ru}, measured $f_h$
with a statistics of 122 events with an identified charmed hadron in
the final state.

After a thorough overview of all available data, we present a method
to extract $f_h$ for neutrino experiments. The major difference of
this method with respect to other presented in the past, i.e.~see
Ref.~\cite{bolton}, is that$f_h$'s from $e^+e^-$ are not extrapolated
to neutrino experiments straightway, but some peculiarities of
$\nu$-induced interactions are accounted for. Indeed, neutrinos may
undergo to quasi-elastic and diffractive processes with production of
charmed hadrons in the final state. These results are also used to
estimate the semi-muonic branching ration of charmed hadrons. Finally,
we discuss the statistical and systematic uncertainties associated
with our predictions and compare them with the E531 results.

\section{Charm-production fraction in DIS interactions}

\subsection{Measurement of the $D^+/D^0$ ratio}

In this Section we give an overview of the available data on $D^+/D^0$
measurements in $e^+e^-$, $\pi N$, $p N$ and $\gamma N$ experiments.
Furthermore, we also study the correlated variable $F_V$, which is
defined as $V/(V+P)$ , where $V$ and $P$ signify vector and
pseudo-scalar charmed mesons, respectively. When needed, we recomputed
$F_V$ by using the latest charmed-hadron branching-ratios
(BR). Finally, we extract for the first time $F_V$ from neutrino induced 
 charm-production data.

In the following we assume that $D^{\star 0}$ and
$D^{\star +}$ production cross-sections are equal at the parton level,
as well as the direct production of $D^0$ and $D^+$.

From here on, we indicate with $D$ the sum of both prompt and decayed
($D^\star\rightarrow D$) $D$ meson production.

Under the previous assumptions, the measurement of $D^+/D^0$ and
$D^\star/D$ ratios allows us to extract $F_V$. By using the formulae
given in Ref.~\cite{na14}, $F_V$ can be extracted from the relations

\begin{equation}
F_V\times B_\star = \frac{1-R_1}{1+R_1}
\label{r1}
\end{equation}

\begin{equation}
F_V\times B_\star = \frac{R_2}{1-R_2}
\label{r2}
\end{equation}

where $R_1\equiv D^+/D^0$, $R_2\equiv (D^0~\mbox{from}~D^{\star +})/D^0$ and
$B_\star=BR(D^{\star +}\rightarrow
D^0\pi^+)=(67.7\pm0.5)\%$~\cite{pdg}.

\subsubsection{$e^+e^-$ experiments}

The basic principle to reconstruct $e^+e^-$ events with charmed
hadrons in the final state is common to all the experiments. Being a
$c\bar{c}$ pair produced in the annihilation, one charmed hadron is
used to tag the events, while the other one to study the decay
properties. The decay modes used to tag the event are~\footnote{Here
and in the following by $D^{+} (D^{*+})$ we implicitly indicate also the
charge conjugated decay modes. }

$$D^{\star +}\rightarrow D^0\pi^+\rightarrow (K^-\pi^+)\pi^+ 
(0.026\pm0.006) $$

$$D^{0}\rightarrow K^-\pi^+ (0.0383\pm0.0009) $$

$$D^{0}\rightarrow K^-\pi^+\pi^+\pi^-  (0.0749\pm0.0031) $$

$$D^{+}\rightarrow K^-\pi^+\pi^+ (0.090\pm0.006 ) $$

where the corresponding branching ratios are also given in
brackets~\cite{pdg}. 

Results on the total cross-sections for inclusive
production of the charmed particles $D^{\star 0}$, $D^{\star +}$,
$D^0$ and $D^+$ at various $\sqrt{s}$ are  shown in
Table~\ref{e+e-low}.

\begin{table}[tbp] 
\begin{center} 
{\small 
\begin{tabular}{||c|c|c|c|c|c|c||} 
\hline 
$\sqrt{s}$ (GeV) & $\sigma{(D^0)}$ (nb) & $\sigma{(D^+)}$ (nb)& $\sigma{(D^{\star 0})}$ (nb) & $\sigma{(D^{\star +})}$ (nb) & $\sigma{(D^+_s)}$ (nb) & $\sigma{(\Lambda^+_c)}$ (nb) \\ 
\hline
\hline
4.03~\cite{bes} & $19.9\pm2.4$ & $6.5\pm0.8$ & ---  & --- & ---  & ---\\
\hline                                                     
4.14~\cite{bes} & $9.3\pm2.4$ & $1.9\pm0.9$ & ---  & ---  & ---  & ---\\
\hline                                                     
4.16~\cite{coles}&$7.8\pm0.8$ & $2.1\pm0.7$ & ---  & ---  & ---  & ---\\
\hline                                                     
5.20~\cite{coles}&$4.7\pm0.8$ & $1.7\pm0.4$ & ---  & ---  & ---  & ---\\
\hline
10.55~\cite{cleo}&$1.36\pm0.16$ & $0.57\pm0.08$ & $0.78\pm0.17$  &  $0.65\pm0.05$ & $0.41\pm0.13$ & $0.20\pm0.08 $\\
\hline
10.55~\cite{argus}&$1.14\pm0.15$ & $0.56\pm0.08$ & --- &  $0.54\pm0.08$ & $0.42\pm0.07$ & $0.45\pm0.07$\\
\hline\hline 
\end{tabular} 
} 
\end{center} 
\caption{Measured cross-sections for the production of charmed hadrons
at center of mass energies ($\sqrt{s}$) in the range $4.03\div
10.55$~GeV.}
\label{e+e-low} 
\end{table}                                                                   

A complete review of the probabilities ($f(c\rightarrow C)$) that a
c-quark fragments into a $D^\star$, $D^0$, $D^+$ and other charmed
hadrons as measured in $Z^0$ decays is given in Ref.~\cite{gladilin}
and reported in Table~\ref{e+e-z0}.

\begin{table}[tbp] 
\begin{center} 
{\small 
\begin{tabular}{||c|c|c|c||c||} 
\hline 
 & ALEPH & DELPHI & OPAL & Weighted average \\
\hline
\hline
$f(c\rightarrow D^0)$(\%) & $55.9\pm1.7\pm1.5$ & $54.5\pm1.5\pm3.2$ & $58.8\pm4.1\pm4.0$ & $55.8\pm1.8$\\ 
\hline
$f(c\rightarrow D^+)$(\%) & $23.8\pm0.8\pm1.3$ & $22.6\pm0.8\pm1.4$ & $23.1\pm3.0\pm2.0$ & $23.2\pm1.0$\\ 
\hline
$f(c\rightarrow D_s)$(\%) & $11.5\pm1.9\pm0.7$ & $12.4\pm1.1\pm1.2$ & $9.0\pm2.4\pm1.1$ &  $11.5\pm1.1$ \\ 
\hline
$f(c\rightarrow \Lambda_c)$(\%) & $ 7.8\pm0.8\pm0.4$ & $ 8.6\pm1.8\pm1.0$ & $ 4.8\pm2.2\pm0.8$ & $ 7.6\pm0.8$ \\ 
\hline
$f(c\rightarrow D^{\star +})$(\%) & $23.3\pm1.0\pm0.8$ & $25.5\pm1.5\pm0.6$ & $22.8\pm0.9$ &  $23.4\pm0.7$ \\ 
\hline\hline 
\end{tabular} 
} 
\end{center} 
\caption{ Measured and averaged probabilities that a charm-quark
  fragments into $D^0$, $D^+$, $D_s$, $\Lambda_c$ and $D^{\star +}$ in
  $e^+e^-$ annihilation at $\sqrt{s}=M_{Z^0}$.}  
\label{e+e-z0} 
\end{table}                                                                   

From Tables~\ref{e+e-low} and~\ref{e+e-z0} we can extract both $R_1$
and $F_V$, the latter being estimated by using Eqs.~(\ref{r1})
and~(\ref{r2}). The results are given in Table~\ref{R1R2} and show
that within the experimental errors, both $R_1$ and $F_V$ are
independent of the energy.

\begin{table}[tbp] 
\begin{center} 
{\small 
\begin{tabular}{||c|c|c|c|c||} 
\hline 
$\sqrt{s}$ (GeV) & $R_1$ & $R_2$ & $F_V(R_1)$ & $F_V(R_2)$ \\ 
\hline
\hline
4.03~\cite{bes} & $0.33\pm0.06$&--- &$0.74\pm0.10$ &--- \\
\hline                          
4.14~\cite{bes} & $0.20\pm0.11$&--- &$0.91\pm0.23$ &---\\
\hline                          
4.16~\cite{coles}&$0.27\pm0.09$&--- &$0.85\pm0.16$ &---\\
\hline                          
5.20~\cite{coles}&$0.36\pm0.10$&--- &$0.70\pm0.16$ &---\\
\hline                          
10.55~\cite{cleo,argus}&$0.45\pm0.06$&$0.32\pm0.04$ &$0.56\pm0.08$ &$0.70\pm0.13$\\
\hline
$M_{Z^0}$& $0.42\pm0.03$& $0.28\pm0.01$  &$0.60\pm0.04$ & $0.57\pm0.03$\\
\hline 
Weighted average & $0.39\pm0.02$& $0.29\pm0.01$ &$0.65\pm0.03$ & $0.60\pm0.03$\\
\hline\hline 
\end{tabular} 
} 
\end{center} 
\caption{Measured $R_1$, $R_2$, $F_V$ in $e^+e^-$ experiments as a function
  of $\sqrt{s}$ and their averaged values.} 
\label{R1R2} 
\end{table}                                                                   

\subsubsection{$\pi N$ and $\gamma N$ experiments}

Several experiments have studied charm-production and extracted $R_1$ and
$F_V$ by using $\pi$ beams of different energies impinging onto
different targets. A non-exhaustive list of all available data in $\pi
N$ experiments is given in Table~\ref{piN}.

The NA14/2 photo-production experiment~\cite{na14} measured both $R_1
= 0.37\pm0.10$ and $R_2 = 0.26\pm0.04$ from which we can extract the
weighted average $F_V = 0.57\pm0.09$. 

From these data we can conclude that, within the experimental errors,
$R_1$ is both process- and energy-independent.

\begin{table}[tbp] 
\begin{center} 
{\small 
\begin{tabular}{||c|c|c|c|c||} 
\hline 
  & $R_1$ & $R_2$  & $F_V(R_1)$ & $F_V(R_2)$\\ 
\hline
\hline
WA92: $350~GeV/c~\pi^-$ & &  & &\\
on Cu, W & $0.423\pm0.012$ & $0.280\pm0.015$  & $0.60\pm0.02$ & $0.57\pm0.04$ \\
\hline
E769: $250~GeV/c~\pi^-$ & &  & &\\
on Be, Al, Cu, W & $0.419\pm0.043$ &  $0.222\pm0.031$ & $0.60\pm0.06$ & $0.42\pm0.08$\\
\hline
E769: $210~GeV/c~\pi^-$ &  &  & &\\
on Be, Al, Cu, W & $0.258\pm0.058$ &--- & $0.87\pm0.11$ &--- \\
\hline
E653: $600~GeV/c~\pi^-$ & &  & &\\
on emulsion & $0.393\pm0.032$ &--- &  $0.64\pm0.05$ &--- \\
\hline
NA32: $230~GeV/c~\pi^-$ & & & & \\
on Cu & $0.422\pm0.033$ &  $0.262\pm0.026$ &$0.60\pm0.05$ & $0.52\pm0.07$\\
\hline
NA32: $200~GeV/c~\pi^-$ & &  & &\\
on Si & $0.439^{+0.123}_{-0.094}$ &  $0.319\pm0.095$ &$0.58\pm0.16$ &$0.69\pm0.30$ \\
\hline
NA27: $360~GeV/c~\pi^-$ & &  & &\\
on H & $0.564\pm0.171$ &---  &$0.41\pm0.21$ &---\\
\hline 
Weighted average & $0.415\pm0.010$ &  $0.268\pm0.012$  &$0.611\pm0.015$ & $0.541\pm0.033$\\
\hline\hline 
\end{tabular} 
} 
\end{center} 
\caption{ Summary of available $R_1$, $R_2$ and $F_V$ measurements
extracted from $\pi$N experiments.}
\label{piN} 
\end{table}                                                                   

\subsubsection{$\nu N$ experiments}
\label{nuexp}

Recently two measurements which allowed us to extract for the first
time $F_V$ from neutrino experiments became available. In
Ref.~\cite{chorus} the CHORUS Collaboration presented a measurement of
the production rate of $D^0$ based on a sample of about $26000$
$\nu_\mu$ charged-current events interactions located and analyzed so
far in the target emulsions. After reconstruction of the event
topology in the vertex region, $283~D^0$ decays were observed with an
estimated background of 9.2 events from $K^0$ and $\Lambda$
decays. The CHORUS Collaboration measured the $D^0$ production
cross-section times $BR(D^0\rightarrow V2)+BR(D^0\rightarrow
V4)]$~\cite{bart}. The total cross-section has been extracted by
accounting for the $D^0$ decays into all neutrals. The value we used
is $BR(D^0\rightarrow~\mbox{all neutral}) =
(25\pm5)\%$~\cite{charles}.

Therefore, the $D^0$ production cross-section normalized to $\nu_\mu$
charged-current (CC) interactions is

\begin{equation}
\frac{\sigma(D^0)}{\sigma_{CC}}=(2.65\pm0.18\pm0.24\pm0.50)\times 10^{-2}
\label{d0}
\end{equation}

at $27~GeV$ average $\nu_\mu$ energy. Notice that this measurement
includes both $D^0$ prompt and $D^0$ from the decay of $D^\star$
mesons. 

The $D^{\star +}$ production in $\nu_\mu$ charged-current interactions
has been measured, with a similar $\nu_\mu$ beam, by the
BEBC~\cite{bebc} and NOMAD~\cite{nomad} experiments to be
$(1.22\pm0.25)\times 10^{-2}$ and $(0.79\pm0.20)\times 10^{-2}$, respectively. The weighted average $D^{\star +}$ production rate normalized to $\nu_\mu$ charged-current interactions is

\begin{equation}
\frac{\sigma(D^{\star +})}{\sigma_{CC}} = (0.96\pm0.16)\times 10^{-2}
\label{dstar}
\end{equation}

From the measured ratios~(\ref{d0}) and~(\ref{dstar}), and by knowing
$B_\star$, we can compute $R_2 = 0.25\pm0.06$. From the latter value
and from Eq.~(\ref{r2}) we can extract

$$F_V = 0.50\pm0.12 $$

It is worth to notice that $R_1$ and $F_V$ extracted from neutrino
experiments can be compared straightway to $e^+e^-$, $\pi N$ and
$\gamma N$ results, being $D^+$ and $D^0$ either produced promptly or
from the decay of prompt $D^{\star +}$ and $D^{\star 0}$. Namely,
processes peculiar of $\nu$ interactions do not affect $R_1$ and
$F_V$~\footnote{ In neutrino interactions $D^{(\star)+}$ may also be
  produced diffractively but, due to the $V_{cd}$ suppression, its
  rate is expected to be about $(1.6\pm0.3)\times 10^{-4}$ with
  respect to CC interactions and therefore negligible.}. 

\subsubsection{Summary and discussion of all available data on $R_1$ and $F_V$}

From results reported in the previous Sections, we can argue that
within the experimental errors $R_1$ is constant over a wide range
of energies ($\sqrt{s}\sim 4 \div 90$~GeV) and independent of the
process. The constant behavior of $R_1$ down to $\sqrt{s}\sim 4$~GeV
can be derived with simple arguments:

\begin{itemize}
\item the masses of $D^+(1869.3)$, $D^0(1864.5)$, $D^{\star
+}(2010.0)$ and $D^{\star 0}(2006.7)$ are very similar. Therefore, the
threshold suppression of $D^\star$ mesons which tends to enhance $D^+$
contribution, is very little;
\item whatever $D^\star/D$ meson is produced a pion should be always
created. Therefore, all charmed mesons have the same threshold
behavior, which cancels out in the ratio.
\end{itemize}

For these reasons we assume that $R_1$ measured in $e^+e^-$ (see
Table~\ref{R1R2}) can be used in neutrino induced charm-production,
too.

\begin{table}[tbp] 
\begin{center} 
{\small 
\begin{tabular}{||c|c|c|c|c|c||} 
\hline 
  & & $F_V$(meas) & $F_V$(UCLA) &  $F_V$(JETSET)& $F_V$(HERWIG)\\ 
\hline
\hline
$e^+e^-$ & $\sqrt{s}=4.03$~GeV & $0.74\pm0.10$ & 0.47 & 0.52& 0.54 \\
\hline 
$e^+e^-$ & $\sqrt{s}=4.14$~GeV & $0.91\pm0.23$ & 0.62 & 0.70& 0.66\\
\hline 
$e^+e^-$ & $\sqrt{s}=4.16$~GeV & $0.85\pm0.16$& 0.63 & 0.71& 0.66\\
\hline 
$e^+e^-$ & $\sqrt{s}=5.20$~GeV & $0.70\pm0.16$& 0.59 & 0.70& 0.28\\
\hline 
$e^+e^-$ & $\sqrt{s}=10.55$~GeV &$0.56\pm0.08$ &0.61 & 0.74& 0.38\\
\hline 
$e^+e^-$ & $\sqrt{s}=91.2$~GeV & $0.60\pm0.04$ & 0.61   & 0.75& 0.39 \\
\hline 
$\pi N$ & $E_\pi=200\div350$~GeV & $0.61\pm0.02$ & --- & --- & --- \\
\hline 
$\gamma N$    &  & $0.57\pm0.09$ & --- & ---  & --- \\
\hline 
$\nu N$       & $E_\nu\sim25$~GeV & $0.50\pm0.12$ & --- & 0.69 & 0.27 \\
\hline\hline 
\end{tabular} 
} 
\end{center} 
\caption{ Summary of the available data on $F_V$ and predictions from
different models. The error for the theoretical predictions is not shown being relevant only for the third digit.}
\label{summary} 
\end{table}                                                                   

As an important by product of our study we have also extracted $F_V$ from
different processes and at several energies (see
Table~\ref{summary}). The simplest model to predict $F_V$ is based on
the spin-counting. Namely, vector mesons are spin-one states
($^3S_1$), while pseudo-scalar mesons are spin-zero states ($^1S_0$),
therefore $F_V=0.75$. The discussion of more refined models (UCLA,
JETSET, HERWIG and others) is beyond the purposes of this paper. For
details we refer to~\cite{Chun:bh}.

From Table~\ref{summary} we can see that the measured $F_V$ is
independent of the processes and of the energy. This means that the
probability for a c-quark to fragment into a $D$ or a $D^\star$ meson
is universal and does not depend neither on the process nor on the
energy. Notice that the UCLA model is the best in describing available
$e^+e^-$ data.

\subsection{Measurement of the $D_s$ to $D^0$ and  $\Lambda_c$ to $D^0$ ratios}

The ratio $D_s/D^0$ has been measured in $e^+e^-$, $\pi N$ and $\gamma
N$ experiments. A summary of the available data is given in
Table~\ref{dstod0}.

The decay mode used to tag the event is

$$D^{+}_s\rightarrow \phi\pi^+\rightarrow (K^-K^+)\pi^+$$

whose BR, as reported by the Particle Data Group, are~\cite{pdg}:

$$BR(D^{+}_s\rightarrow \phi\pi^+) = 0.036\pm0.009 $$
$$BR(\phi\rightarrow K^-K^+) = 0.492\pm0.007 $$

From Table~\ref{dstod0} we can see that the $D_s/D^0$ ratio is, within
the errors, independent of the energy and of the process. 

Data on the ratio $\Lambda_c/D^0$ are very poor. Indeed, it has been
measured only in $e^+e^-$ experiments (see Table~\ref{dstod0}).

The decay mode used to tag the event is

$$\Lambda_c\rightarrow p K^-\pi^+$$

whose BR, as reported by the Particle Data Group, is~\cite{pdg}:

$$BR(\Lambda_c\rightarrow p K^-\pi^+) = 0.050\pm0.013 $$

From Table~\ref{dstod0} we can see that, although with a smaller
statistical accuracy, both $R_s$ and $R_{c}$ are, within the
errors, independent of the energy. 

\begin{table}[tbp] 
\begin{center} 
{\small 
\begin{tabular}{||c|c|c||} 
\hline 
$\sqrt{s}$ (GeV) & $R_s\equiv D_s/D^0$ & $R_{c}\equiv \Lambda_c/D^0$ \\ 
\hline
\hline
4.14~\cite{bes} & $0.176\pm0.076$ & --- \\
\hline                                                            
10.55~\cite{cleo}&$0.148\pm0.052$ & $0.148\pm0.052$ \\
\hline                                                            
10.55~\cite{argus}&$0.184\pm0.040$& $0.158\pm0.039$ \\
\hline                                                            
$M_{Z^0}$ (ALEPH) & $0.206\pm0.039$           &  $0.140\pm0.017$\\
\hline                                                            
$M_{Z^0}$ (DELPHI) & $0.228\pm0.033$          &  $0.158\pm0.039$\\
\hline                                                            
$M_{Z^0}$ (OPAL) & $0.153\pm0.047$            &  $0.082\pm0.041$ \\
\hline 
WA92: $350~GeV/c~\pi^-$ &         & --- \\
on Cu& $0.160\pm0.037$            & \\
\hline
WA92: $350~GeV/c~\pi^-$ &         & --- \\ 
on W& $0.183\pm0.068$             & \\
\hline
NA14/2 ($\gamma N$)& $0.185\pm0.083$           &  --- \\
\hline\hline 
Weighted average & $0.193\pm0.016$&  $0.135\pm0.015$\\
\hline\hline 
\end{tabular} 
} 
\end{center} 
\caption{ Summary of the available $R_s$ and $R_c$ measurements
extracted from $e^+e^-$, $\nu$N and $\gamma$N experiments. The $\pi N$ data have been taken from Refs.~\cite{beatrice,beatrice2} and references therein, while $\gamma N$ results from Ref.~\cite{na14}. }
\label{dstod0} 
\end{table}                                                                   

\subsubsection{Summary and discussion of all available data on $R_s$ and $R_{c}$}
Although from Table~\ref{dstod0} it seems that $R_s$ is constant over
a wide range of energies ($\sqrt{s}\sim 4 \div 90$~GeV) and
independent of the process, some comments are in order.

Given the quark composition of the $D_s^+(c\bar{s})$ meson, it has to
be created always together with at least one $K$ meson. Therefore,
being $m_K\approx 500$~MeV, we expect that the threshold effect for
$D_s$ production is more pronounced than for $D$ mesons, when at least
one $\pi$ has to be produced ($m_\pi\approx 100$~MeV). To account for
the different threshold effect at low energies, in the following we do
not use the $R_s$ values measured at the $Z^0$ peak. Furthermore,
under the assumption that $\sqrt{s}$ in collider experiments can be
replaced with $ W$ in fixed target experiments, $W$ being the final
state hadronic mass, and noting that neutrino-induced charm events at 
present experiments are characterized by values of $W$ in the range
$4\div 10$~GeV, we can argue a $R_s$ value of $R_s=0.171\pm0.029$
which corresponds to the weighted average of the first three results
in Table~\ref{dstod0}. 

The available measurements on $R_c$ are very poor. Nevertheless, as
it will be discussed in Section~\ref{neutrino}, we do not use the
$R_c$ value to predict the charmed fractions in events induced by
neutrinos.

\section{Charm-production fractions in neutrino experiments}
\label{neutrino}

\subsection{The method}
\label{method}
From the previous Sections we can argue that once a charm-quark has
been produced in deep-inelastic interactions (i.e.~the energy of the
process is higher than the threshold), it has a probability to produce
a charmed hadron $C_h$ which is, within the experimental errors,
independent of the process and of the energy. Therefore, as far as
the deep-inelastic scattering (DIS) is concerned, we can write the charm
production rate as

\begin{eqnarray}
\frac{\sigma_c(DIS)}{\sigma_{CC}} & = &
 \frac{\sigma_{D^0}}{\sigma_{CC}} + \frac{\sigma_{D^+}}{\sigma_{CC}} +
 \frac{\sigma_{D_s}}{\sigma_{CC}} +
 \frac{\sigma_{\Lambda_c}}{\sigma_{CC}} \nonumber \\ & = &
 \frac{\sigma_{D^0}}{\sigma_{CC}}\times\left(
 1+R_1+R_s+R_{c}\right)\, .
\label{DIS}
\end{eqnarray}

In the case of neutrino induced charm-production, Eq.~(\ref{DIS}) is
not correct. Indeed, in this case we also have to account for
diffractive and quasi-elastic charm-production. Therefore, the
inclusive charm-production rate in neutrino interactions can be
written as

\begin{equation}
\frac{\sigma_c(\nu)}{\sigma_{CC}} =  \frac{\sigma_c(DIS)}{\sigma_{CC}}
+ \frac{\sigma_{D_s}}{\sigma_{CC}}\mid_{diff} +
\frac{\sigma_{\Lambda_c}}{\sigma_{CC}}\mid_{QE} \, . 
\end{equation}

If one accepts that $R_1$, $R_s$ and $R_{c}$ from $e^+e^-$ (with
$\sqrt{s}=4.1\div90$~GeV) and other experiments can be used to
described the fragmentation of charm-quarks produced in DIS neutrino
interactions with average final state hadronic mass $\langle
W\rangle\sim 10$~GeV, then

\begin{eqnarray}
\frac{\sigma_c(\nu)}{\sigma_{CC}} & = &
 \frac{\sigma_{D^0}}{\sigma_{CC}}\times\left(
 1+R_1+R_s+R_{c}\right) +
 \frac{\sigma_{D_s}}{\sigma_{CC}}\mid_{diff} +
 \frac{\sigma_{c}}{\sigma_{CC}}\mid_{QE}
\label{neuall}
\end{eqnarray}

From Eq.~(\ref{neuall}) charm-production fractions in neutrino
interactions can be written as

$$f_{D^0}(E_\nu) =  \frac{\sigma_{D^0}}{\sigma_{CC}}(E_\nu)\times\frac{1}{\frac{\sigma_c(\nu)}{\sigma_{CC}}(E_\nu)} $$

$$ f_{D^+}(E_\nu) =  R_1\times\frac{\sigma_{D^0}}{\sigma_{CC}}(E_\nu)\times\frac{1}{\frac{\sigma_c(\nu)}{\sigma_{CC}}(E_\nu)} $$

$$ f_{D^+_s}(E_\nu) =  \left( R_s\times\frac{\sigma_{D^0}}{\sigma_{CC}}(E_\nu) + \frac{\sigma_{D_s}}{\sigma_{CC}}\mid_{diff}(E_\nu)\right )\times\frac{1}{\frac{\sigma_c(\nu)}{\sigma_{CC}}(E_\nu)} $$

Notice that, given the poor knowledge on $R_{c}$
and on the quasi-elastic charm-production cross-section, we derive
$f_{\Lambda_c}$ by using the normalization constrain
$f_{D^0}+f_{D^+}+f_{D^+_s}+f_{\Lambda^+_c}=1$. 
In order to estimate the charmed fractions and their energy dependence, we use the following inputs

\begin{itemize}
\item the inclusive charm-production rate derived in Ref.~\cite{crosscharm};
\item the energy dependence of the $D^0$ production rate reported in
Ref.~\cite{chorus} properly scaled to account for the effect discussed
in Section~\ref{nuexp};
\item the energy dependence of the diffractive $D_s$ production given
in~\cite{ds}, properly scaled in order to reproduce the average
diffractive charm-production cross-section as measured by BEBC and
NuTeV~\cite{fds} ($\frac{\sigma_{D_s}}{\sigma_{CC}}\mid_{diff}=
(0.31\pm0.05)\times10^{-2}$).
\end{itemize}

\subsection{Results on $f_h$ and $B_\mu$ and comparison with the data}

\begin{figure}[hbtp]
  \begin{center} \resizebox{0.8\textwidth}{!}{
    \includegraphics{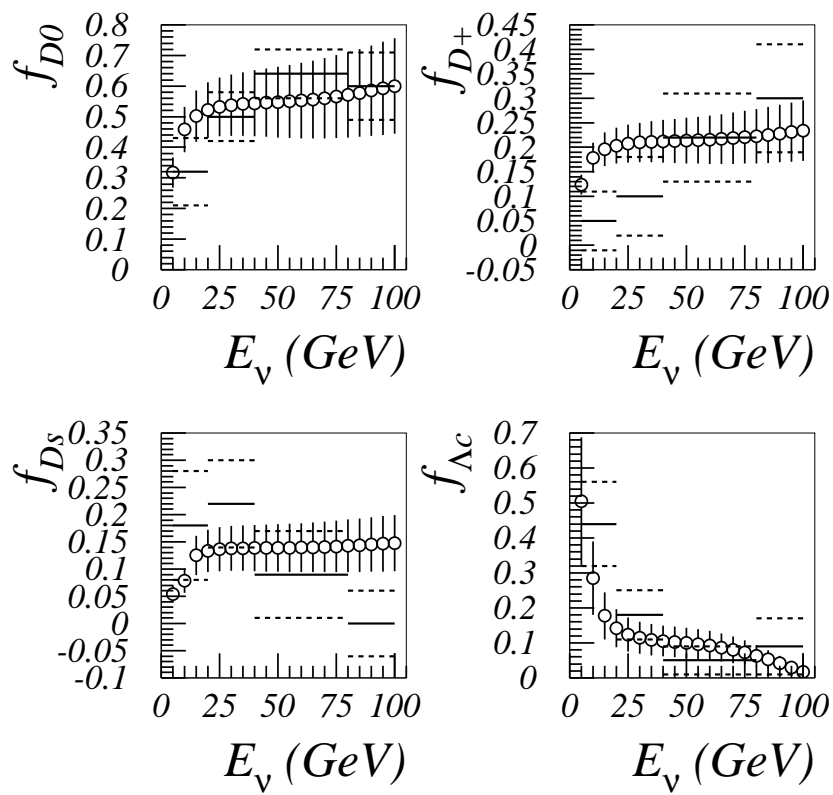} }\caption{\small Predicted
    charm-production fractions as a function of the neutrino energy. The
    continous line shows the central value of the $f_h$ as measured by
    E531, while the dashed lines the upper and lower bounds at 68\%
    CL.}  \label{fig:charfrac} \end{center}
\end{figure}

By using the method described in the previous section, we derived the
charm-production fractions, as a function of the neutrino energy, as
reported in Table~\ref{chfrac} and shown in
Fig.~\ref{fig:charfrac}. Our results are in good agreement with the
charm-production fractions extracted from the E531 data, see
Fig.~\ref{fig:charfrac}. It is worth noticing that $f_{\Lambda^+_c}$
shows a dependence on the energy, higher values at low neutrino
energies, consistent with the expectations. Indeed, quasi-elastic
charm-production, which yields only $\Lambda_c$, is expected to
largely contribute to $\sigma_c(\nu)$ for
$E_\nu<25$~GeV~\cite{bolton}. In Table~\ref{chfrac} the expected
charm-production fractions in the CHORUS experiment are also given.

\begin{table}[tbp] 
\begin{center} 
{\small 
\begin{tabular}{||c|c|c|c|c||} 
\hline 
$E_\nu$ (GeV) & $f_{D^0}$ & $f_{D^+}$  & $f_{D^+_s}$ & $f_{\Lambda^+_c}$\\ 
\hline
\hline
5 &$0.32\pm0.05$ &$0.12\pm0.02$ &$0.054\pm0.015$ & $0.50\pm0.18$\\
\hline
10 &$0.46\pm0.07$ &$0.18\pm0.03$ &$0.078\pm0.022$ & $0.29\pm0.11$\\
\hline
15 &$0.50\pm0.08$ &$0.20\pm0.03$ &$0.13\pm0.04$ & $0.18\pm0.07$\\
\hline
20 &$0.52\pm0.09$ &$0.20\pm0.04$ &$0.13\pm0.04$ & $0.14\pm0.05$\\
\hline
25 &$0.53\pm0.10$ &$0.21\pm0.04$ &$0.14\pm0.04$ & $0.12\pm0.05$\\
\hline
30 &$0.54\pm0.10$ &$0.21\pm0.04$ &$0.14\pm0.04$ & $0.11\pm0.05$\\
\hline
35 &$0.54\pm0.10$ &$0.21\pm0.04$ &$0.14\pm0.04$ & $0.11\pm0.04$\\
\hline
40 &$0.54\pm0.11$ &$0.21\pm0.04$ &$0.14\pm0.04$ & $0.11\pm0.04$\\
\hline
50 &$0.55\pm0.12$ &$0.21\pm0.05$ &$0.14\pm0.04$ & $0.10\pm0.04$\\
\hline
60 &$0.55\pm0.12$ &$0.22\pm0.05$ &$0.14\pm0.04$ & $0.09\pm0.04$\\
\hline
70 &$0.56\pm0.13$ &$0.22\pm0.05$ &$0.14\pm0.04$ & $0.08\pm0.04$\\
\hline
80 &$0.57\pm0.14$ &$0.22\pm0.06$ &$0.14\pm0.04$ & $0.06\pm0.03$\\
\hline
90 &$0.58\pm0.15$ &$0.23\pm0.06$ &$0.15\pm0.04$ & $0.04\pm0.02$\\
\hline
100 &$0.60\pm0.16$ &$0.23\pm0.06$ &$0.15\pm0.05$ & $0.02\pm0.01$ \\
\hline
CHORUS    &$0.524\pm0.036$ &$0.204\pm0.014$ &$0.128\pm0.012$ &$0.147\pm0.008$ \\
\hline\hline 
\end{tabular} 
} 
\end{center} 
\caption{Prediction of charm-production fractions in neutrino induced
  events as a function of the neutrino energy.} 
\label{chfrac} 
\end{table}                                                                   

Having determined the $f_h$'s, we can also estimate the semi-muonic
branching ratio $B_\mu$ of the charmed hadrons as a function of the
neutrino energy.  $B_\mu$ is a very important quantity, being the
input variable needed to extract from the dimuon data the element of
the CKM matrix $V_{cd}$.

Recently, a direct measurement of $\bar{B}_\mu$ has been performed by
the CHORUS Collaboration by using a statistics of about 1000 charm
events reconstructed in the nuclear emulsions. Out of these,
($88\pm10\pm8$) dimuon events have been reconstructed, which
correspond to~\cite{bart}

$$\bar{B}_\mu = (9.3\pm1.3)\%\, .$$

This measurement has to be compared with our prediction obtained by
convoluting the charm-production fractions with the CHORUS neutrino flux

$$\bar{B}_\mu = (8.8\pm1.0)\%\, .$$

\begin{table}[tbp]
\begin{center}
{\small
\begin{tabular}{|c|c|c|c|}
\hline
Collaboration & $E_\nu$~(GeV) & $B_\mu\mid V_{cd}\mid^2$ & $B_\mu$ \\
\hline \hline
CDHS~\cite{Abramowicz:1982zr} & 20.0 & $(0.41\pm0.07)\times 10^{-2}$ & $0.083\pm0.014$ \\
\hline
CHARM II~\cite{Vilain:1998uw} &23.6 & $(0.442\pm0.049)\times 10^{-2}$ & $0.090\pm0.010$ \\
\hline
NOMAD~\cite{Astier:2000us}   & 23.6  &$(0.48\pm0.17)\times 10^{-2}$ &  $0.097\pm0.034$ \\
\hline
CHORUS~\cite{bart}   & 27  & &  $0.093\pm0.013$ \\
\hline
CCFR (LO)~\cite{Rabinowitz:1993xx} & 140   & $(0.509\pm0.036)\times 10^{-2}$ & $0.103\pm0.007$  \\
\hline
CCFR (NLO)~\cite{Bazarko:1994tt} & 140  & $(0.534\pm0.060)\times 10^{-2}$ & $0.108\pm0.012$ \\
\hline
\hline
{\bf PDG~\cite{pdg}}& $V_{cd}=0.219\div0.225$ & $\langle V_{cd}\rangle=0.222\pm0.003$ & \\
(From unitarity at 90\%)& & & \\
\hline
\hline
\end{tabular}}
\end{center}
\caption{Charmed hadron semi-muonic branching ratios and $V_{cd}$
measured by various experiments at different neutrino energies. The
direct measurement performed by CHORUS is also shown.}
\label{tab:dimuon}
\end{table}

In Table~\ref{tab:dimuon} we derived, by using the $V_{cd}$ value
obtained by imposing the unitarity constraint to the CKM matrix and
the measurements of $B_\mu\mid V_{cd}\mid^2$ from various experiments,
$B_\mu$. Given the fact that the different experiments exploit
different neutrino energy spectra, we can probe the sensitivity of
$B_\mu$ to the neutrino energy. As expected, the higher the neutrino
beam energy the larger the value of $B_\mu$.

%{\bf NB The charmed fractions measured in CHORUS and shown in
%  Table~\ref{chfrac} are still very preliminary and for internal use only!}

%For anti-neutrino interactions the charm-production cross-section can be
%written as

%\begin{eqnarray}
%\frac{\sigma_c(\bar{\nu})}{\sigma_{CC}} & = &
% \frac{\bar{\sigma}_{D^0}}{\bar{\sigma}_{CC}}\times\left(
% 1+R_1+R_s+R_{c}\right) +
% \frac{\bar{\sigma}_{D^s}}{\bar{\sigma}_{CC}}\mid_{diff}
%\label{antineuall}
%\end{eqnarray}

%From Eq.~\ref{neuall} we can write the charm-production fractions in
%neutrino interactions as
%
%$$\bar{f}_{D^0} =  \frac{\bar{\sigma}_{D^0}}{\bar{\sigma}_{CC}}\times\frac{1}{\frac{\sigma_c(\bar{\nu})}{\bar{\sigma}_{CC}}} $$
%
%$$ \bar{f}_{D^+} =  R_+\times\frac{\bar{\sigma}_{D^0}}{\bar{\sigma}_{CC}}\times\frac{1}{\frac{\sigma_c(\bar{\nu})}{\bar{\sigma}_{CC}}} $$
%
%$$ \bar{f}_{D^+_s} =  \left( R_s\times\frac{\bar{\sigma}_{D^0}}{\bar{\sigma}_{CC}} + \frac{\bar{\sigma}_{D^s}}{\bar{\sigma}_{CC}}\mid_{diff}\right )\times\frac{1}{\frac{\sigma_c(\bar{\nu})}{\bar{\sigma}_{CC}}} $$
%
%$$ \bar{f}_{\Lambda^+_c} = R_{\Lambda_c^+}\times\frac{\bar{\sigma}_{D^0}}{\bar{\sigma}_{CC}}\, . $$
%
%
%%%%%%%%%%%%%%%%%%%%%%%%%%%%%%%%%%%%%%%%%%%%%%%%%%%%%%%%%%%%%%%%%%%%%%%%%%%%%%%
% Conclusions
%%%%%%%%%%%%%%%%%%%%%%%%%%%%%%%%%%%%%%%%%%%%%%%%%%%%%%%%%%%%%%%%%%%%%%%%%%%%%%
%

\section{Conclusions}
We have presented a method to extract the charmed fractions in
neutrino induced events. The method relies on the fact that, apart
from processes peculiar to neutrinos such as quasi-elastic and
diffractive charm-production, the charm-production and fragmentation
mechanism is believed to be process-independent. We have verified this
natural assumption going through a complete review of the available
data from different experiments. Moreover, the $D^+$ over $D^0$ ratio
is constant over the large energy range spanned by the collider and
fixed target experiments reviewed in this paper. By using recent data
from neutrino experiments, we have assessed the consistency of this
ratio with the one predicted by other experiments. On the other side,
the energy-independent behavior of the ratio itself is clear from the
review of the experiments. 

Threshold effects for the $D_s$ production are seen and accounted for.
By introducing the diffractive charm-production and using the unity
constraint, we have predicted the charm fragmentation as a function of
the neutrino energy in the range useful to present neutrino
experiments. In particular, a prediction for the CHORUS experiment has
been made. The determination of the fragmentation function also allows
the prediction of the semi-muonic branching ratio of charmed
hadrons. The prediction given is in good agreement with a recent
measurement made by the CHORUS experiment.

%
%%%%%%%%%%%%%%%%%%%%%%%%%%%%%%%%%%%%%%%%%%%%%%%%%%%%%%%%%%%%%%%%%%%%%%%%%%%%%%%
% Acknowledgments
%%%%%%%%%%%%%%%%%%%%%%%%%%%%%%%%%%%%%%%%%%%%%%%%%%%%%%%%%%%%%%%%%%%%%%%%%%%%%%
%
%\section*{Acknowledgments}

%
%%%%%%%%%%%%%%%%%%%%%%%%%%%%%%%%%%%%%%%%%%%%%%%%%%%%%%%%%%%%%%%%%%%%%%%%%%%%%%%
% Bibliography
%%%%%%%%%%%%%%%%%%%%%%%%%%%%%%%%%%%%%%%%%%%%%%%%%%%%%%%%%%%%%%%%%%%%%%%%%%%%%%
%

\newpage
%
%%%%%%%%%%%%%%%%%%%%%%%%%%%%%%%%%%%%%%%%%%%%%%%%%%%%%%%%%%%%%%%%%%%%%%%%%%%%%%%
% Figures
%%%%%%%%%%%%%%%%%%%%%%%%%%%%%%%%%%%%%%%%%%%%%%%%%%%%%%%%%%%%%%%%%%%%%%%%%%%%%%%
%

%%%%%%%%%%%%%%%%%%%%%%%%%%%%%%%%%%%%%%%%%%%%%%%%%%%%%%%%%%%%%%%%%%%%%%%%%%%%%%%
\end{document}